\documentstyle[12pt,aaspp4]{article}
\begin{document}
\parindent=1.0cm

\title{The M31 Dwarf Spheroidal Companion And V: $g', r',$ and $i'$ Imaging 
with GMOS on Gemini North\footnote[1]{Based on observations obtained at the 
Gemini Observatory, which is operated by the Association of Universities 
for Research in Astronomy, Inc., under a co-operative agreement with the 
NSF on behalf of the Gemini partnership: the National Science Foundation 
(United States), the Particle Physics and Astronomy Research Council 
(United Kingdom), the National Research Council of Canada (Canada), 
CONICYT (Chile), the Australian Research Council (Australia), CNPq (Brazil), 
and CONICET (Argentina).}}

\author{T. J. Davidge}

\affil{Canadian Gemini Office, Herzberg Institute of Astrophysics,
\\National Research Council of Canada, 5071 W. Saanich Road,\\Victoria,
B.C. Canada V9E 2E7\\ {\it email:tim.davidge@nrc.ca}}

\author{G. S. Da Costa}

\affil{Research School of Astronomy and Astrophysics,\\Australian National 
University, Mount Stromlo Observatory,\\Cotter Road, Weston ACT 2611, 
Australia\\ {\it email: gdc@mso.anu.edu.au}}

\author{Inger J{\o}rgensen}

\affil{Gemini Observatory, Northern Operations Center,\\670 N. A'ohoku Place, 
Hilo, Hawaii, 96720 USA\\ {\it email:ijorgensen@gemini.edu}}

\author{J. R. Allington-Smith}

\affil{Department of Physics, Durham University,\\Rochester Building, 
Science Laboratories, South Road,\\Durham, UK DH1 3LE \\ 
{\it email: j.r.allington-smith@durham.ac.uk}}

\begin{abstract}

	Images obtained in $g', r'$, and $i'$ with the Gemini 
Multi-Object Spectrograph (GMOS) on Gemini North are used to investigate the 
metallicity and stellar content of the M31 dwarf spheroidal companion galaxy 
And V\@. Red giant branch (RGB) stars are traced out to radii in excess of 
$126\arcsec$ from the galaxy center, indicating that And V extends over a 
diameter approaching 1 kpc. The mean $(g'-i')$ color of the RGB does not 
change with radius. Based on the slope of the RGB in the $(i', g'-i')$ 
color-magnitude diagram (CMD), we conclude that the metallicity of And V 
is [Fe/H] = $-2.2 \pm 0.1$. This is lower than earlier estimates, and places 
And V squarely on the relation between metallicity and integrated 
brightness defined by other dwarf spheroidal and dwarf 
elliptical galaxies. In contrast to many of the Galaxy's dwarf 
spheroidal companions, there is no evidence for a statistically significant 
population of luminous asymptotic giant branch stars near the center of 
the galaxy.

\end{abstract}


\section{INTRODUCTION}

	There is a growing body of evidence that galaxies form in a hierarchal 
manner, with low mass systems serving as the basic building 
blocks of galaxy formation. Not only are hierarchal galaxy formation models 
able to reproduce a wide range of the observational properties of nearby 
galaxies (e.g. Somerville \& Primack 1999; Cole et al.\ 2000), but they also 
provide a natural explanation for extended stellar streams 
(e.g. Ibata et al. 2001; Martinez-Delgado et al. 2001) and 
the overall structural properties of the Galactic halo (Bullock, 
Kravtsov, \& Weinberg 2001). The merging process 
continues to the present day, and it has been suggested that 
the ultimate fate of the Local Group may be a merger that produces an 
elliptical galaxy (Forbes et al.\ 2000). 

	Studies of present-day dwarf galaxies can provide 
insights into the systems from which larger galaxies formed. For example, 
there are indications that the Galactic halo could not have formed from 
the accretion of systems with chemical enrichment histories similar to those 
of the low mass dwarf spheroidal companions of the Milky-Way (Shetrone, 
C\^{o}t\'{e}, \& Sargent 2001). One possible explanation is that the 
chemical evolution of the present-day dwarf spheroidal 
companions of the Milky-Way has been affected by tidal interactions 
with the Galaxy (e.g. Mayer et al. 2001), with the result that these 
systems evolved differently from those that 
merged early-on to form the Galactic halo.

	CDM-based simulations predict 
many more low-mass systems in the Local Group than have been detected 
(e.g. Kauffmann, White, \& Guiderdoni 1993; Klypin et al.\ 1999), 
and this provides a strong motivation to search for 
heretofore undiscovered dwarf galaxies. Armandroff et al.\ (1998) 
discuss such a search for low surface brightness companions of M31, and report 
the detection of the dwarf galaxy And V\@. Using a color-magnitude 
diagram (CMD) obtained from follow-up CCD images, Armandroff 
et al.\ (1998) conclude that the metallicity and mean central 
surface brightness of And V are superficially unremarkable, in the sense that 
these quantities are similar to those of two other companions of M31: And I 
and And III.

	The apparent similarity to And I and And III notwithstanding, with a 
mean abundance [Fe/H] $= -1.5$ and absolute magnitude M$_V = -9$, And V falls 
well off of the relation between metallicity and M$_V$ that is otherwise 
well-defined by observational data (e.g.\ Caldwell 1999). The [Fe/H] -- 
M$_V$ relation is generally thought to be imprinted during the early
stages of galaxy evolution, and is a measure of the extent of chemical 
enrichment before the onset of winds that purge a galaxy of its interstellar 
medium. This relation is of fundamental astrophysical importance; for example, 
it is likely the physical basis behind the 
color-magnitude relation of early-type galaxies (Kodama \& Arimoto 1997), 
and is a basic prediction of CDM-dominated galaxy formation models 
(e.g.\ Dekel \& Silk 1986, Cole et al.\ 2000). 

	Why does And V depart from the [Fe/H] -- M$_V$ relation? A 
reasonable starting point for answering this question is to check the existing 
metallicity determination. The CCD data used by Armandroff et al.\ (1998) to 
construct the CMD of And V were recorded during $1\arcsec$ seeing conditions, 
and are restricted to the upper $\sim 1.5$ mag of the red giant branch (RGB). 
Their CMD shows a scatter of $\pm 0.1$ mag, and includes only a 
fraction of the stars in the galaxy. Given the potential significance of And V, 
we decided to re-investigate the metallicity of this galaxy using 
(1) deep multi-color images recorded during sub-arcsec seeing conditions and 
(2) spectra of individual stars that include the near-infrared Ca 
II triplet, which is a metallicity indicator (e.g.\ Armandroff \& Da Costa 
1991). In the present paper we discuss the photometric measurements of stars in 
And V; the spectroscopic observations will be discussed in a second 
paper (Da Costa et al.\ in preparation). Based on the slope of the giant 
branch in the CMD, we find that [Fe/H] $= -2.2\pm 0.1$, indicating that 
And V is more metal-poor than previously thought.

\section{OBSERVATIONS AND REDUCTIONS}

	The data were obtained at the Gemini North telescope 
as part of the System Verification program for 
the Gemini Multi-Object Spectrograph (GMOS). Detailed descriptions of GMOS 
have been provided by Crampton et al.\ (2000) and Davies et al.\ (1997), and 
so we restrict the present discussion to the GMOS detector. 
The detector for GMOS is a mosaic of three $2048 \times 4608$ EEV CCDs, 
each pixel of which subtends $0\farcs0727$ on a side. A $5\farcm5$ diameter 
field is imaged on the detector array, which is over-sized 
so that spectra of objects near the field edge can be obtained. The 
detector was binned $2 \times 2$ at the telescope for these observations, so 
each supra-pixel subtends $0\farcs145$ on a side.

	Images of And V were recorded through $g', r'$, and $i'$ filters 
(Fukugita et al.\ 1996). A series of nine exposures were obtained in each 
filter, and the telescope was offset between these to allow the gaps between 
the individual CCDs to be filled in when constructing the final images. 
The total exposure times and FWHMs of the And V data are listed in Table 1.

	The central regions of the globular clusters NGC 288, NGC 2419, and 
M2 (NGC 7089) were also imaged in $g'$ and $i'$ to calibrate the metallicity 
dependence of $g'-i'$; $r'$ images were also taken of NGC 2419. The exposure 
times and FWHM of the cluster data are also listed in Table 1, while the 
metallicities of these clusters, as given in the current version (June 1999) 
of the Harris (1996) database, are listed in Table 2.

	The photometric calibration was defined using observations of 
15 Landolt (1992) standard stars that were obtained over 3 
different nights during the October system verification observing run. 
The standard magnitudes were transformed into the $g'r'i'$ system using 
the relations given in Fukugita et al. (1996). Mean atmospheric 
extinction coefficients for Mauna Kea were used, and the 
estimated uncertainties in the zeropoints are $\pm 0.05$ mag. Conditions 
were photometric when the data discussed in this paper were recorded.

	The data were reduced using the processing pipeline developed 
for GMOS and implemented in the Gemini IRAF package. 
The reduction steps are (1) bias subtraction, 
(2) flat-field correction, including the correction of CCD-to-CCD gain 
differences, (3) the construction of a mosaic from the individual 
CCD images, taking into account the spatial shifts and rotations between 
the detector elements, (4) the spatial registration of the mosaiced images, 
and (5) co-addition of the mosaiced images. The latter step takes into 
account offsets made at the telescope to fill in the gaps between the 
CCDs, and also cleans the images of bad pixels and 
cosmic ray hits. The results were then trimmed to the area of common coverage, 
so that the exposure time was constant across the frame. 
The central portion of the final $i'$ image, which 
contains the main body of And V, is shown in Figure 1.

\section{RESULTS}

\subsection{Photometric Measurements}

	Stellar brightnesses were measured with the PSF-fitting program 
ALLSTAR (Stetson \& Harris 1988), with PSFs and preliminary 
aperture measurements obtained from tasks in DAOPHOT (Stetson 1987).
Completeness fractions and internal uncertainties in the And V 
measurements were estimated from artificial star experiments in which 
scaled versions of the PSF, with brightnesses and colors tracing the locus 
of actual stars in the And V CMDs, were added to the final images. Photometric 
measurements of the artificial stars were made using the same procedures 
employed for real stars. The predicted completeness fractions and internal 
errors in the photometry in $g'$ and $i'$ are shown in Figure 2.

	The data reduction process included shifting and rotating individual 
images, and this could potentially affect the photometry. To estimate the 
extent of any systematic effects introduced by this processing, 
stellar brightnesses were measured from a portion of a single bias-subtracted 
and flat-fielded $i'$ image that included the center of And V, and the results 
were compared with the photometry from the final co-added $i'$ image. The two 
sets of measurements are in excellent agreement when $i' < 21$, which is where 
the difference in integration times has the least impact on the photometry. The 
mean difference in brightness for 25 stars having $i' < 21$ is 0.002 mag with a 
standard deviation of $\pm 0.010$ mag. For the 20 stars with $i'$ between 21.0 
and 21.5, which is the top 0.5 mag of the And V RGB, the difference is 0.010 
mag with a standard deviation $\pm 0.026$ mag. After adjusting for the lower 
integration time of the single image, the standard deviation is 
consistent with the photometric scatter predicted from the artificial 
star experiments. Therefore, we conclude that the processing used to construct 
the final mosaiced co-added images has not affected the photometry.

	And V is located at an intermediate Galactic latitude, and there is 
significant contamination from field stars and 
background galaxies at the brightnesses corresponding 
to giants in this galaxy. Given that And V is relatively compact and round 
(Caldwell 1999), and that the fractional contamination from field objects 
depends on distance from the galaxy center, we decided to investigate the 
photometric properties of sources in 3 equal-area annuli centered on And V. The 
$(i', g'-i')$ CMDs of stars in these annuli are plotted in Figure 3; 
the radial dimensions used to define each annulus are also 
listed in this figure. Caldwell (1999) traced the light profile of And V out to 
a radius of $\sim 120\arcsec$. However, the And V giant branch can be detected 
even in the outermost annulus data plotted in Figure 3, indicating that 
And V extends to radii in excess of those explored by Caldwell (1999); 
consequently, the integrated brightness of And V estimated by Caldwell (1999) 
is likely a lower limit.

	If mean metallicity and/or age varies in a systematic way with radius 
in And V then this will affect the location 
of the RGB in the CMDs. To determine if there is 
evidence for a population gradient, we measured the mean $g'-i'$ color 
between $i'=22.75$ and $i'=23.25$, and found that $\overline{g'-i'} = 1.100 
\pm 0.009, 1.112 \pm 0.015,$ and $1.076 \pm 0.032$ for the inner, middle, 
and outer annuli. The giant branch is not vertical, and annulus-to-annulus 
differences in the distribution of points along the $i'$ axis could affect the 
means computed above. Therefore, we also computed the mean of the quantity 
$b = (g'-i')-\frac{\Delta(g'-i')}{\Delta(i')} \times (i' - 23)$ in each 
annulus for stars between $i' = 22.75$ and $i'=23.25$. A least squares fit to 
the inner annulus RGB indicates that the RGB slope at this brightness is 
$\frac{\Delta(g'-i')}{\Delta(i')} = -0.261 \pm 0.001$, and we adopted this 
value to compute $b$. We find that $\overline{b} = 1.106 \pm 0.008$, $1.108 \pm 
0.015$, and $1.098 \pm 0.028$ in the inner, middle, and outer annuli. We 
conclude that the color of the giant branch does not change significantly with 
distance from the center of And V, and so there is no evidence 
for a population gradient. This is consistent with the results of other 
M31 dwarf spheroidals (e.g.\ Da Costa, Armandroff, \& Caldwell 2002). Indeed, 
in no dwarf spheroidal system has any radial dependence of the RGB 
color been detected, with the possible exception of Fornax (Saviane, Held, 
\& Bertelli 2000). 

	The cluster CMDs were corrected for reddening and distance using the 
$B-V$ color excesses and $V$ HB brightnesses listed in the current version 
(June 1999) of Harris (1996). Distances were based on the Salaris \& Cassisi 
(1997) HB calibration; the reasons for adopting this particular calibration are 
discussed in \S 3.2. The resulting $(M_{i'}, (g'-i')_0)$ CMDs of the globular 
clusters are plotted in Figure 4. 

	A crowded central field was intentionally imaged in each 
cluster in an effort to sample a significant number of bright giants for 
subsequent spectroscopic follow-up; nevertheless, despite the relatively 
high stellar densities, the CMDs are well-defined and 
extend below the main sequence turn-off.
The scatter in the CMDs indicate that the internal 
photometric accuracy is on the order of a few hundredths of a mag. The 
NGC 288 CMD is the best-defined, as the field observed for this 
cluster has the lowest stellar density, and hence is least affected by 
crowding. However, the giant branch of this cluster in Figure 4 
terminates $0.5 - 1.0$ mag fainter than that of the other 2 clusters. 
We note that the light profiles presented by Trager, King, \& 
Djorgovski (1995) indicate that the central surface brightness of NGC 288 is 
roughly 5 mag/arcsec${^2}$ fainter in M$_V$ than for the other two clusters.
The correspondingly lower central stellar density then explains the lack of 
bright giants: the number of stars in the NGC 288 field is simply not high 
enough to permit stars on the uppermost regions of the RGB to be observed. 

\subsection{The Metallicity of And V}

	To facilitate comparisons between the And V and globular cluster giant 
branches, normal points were computed from the $(i',g'-i')$ cluster CMDs. 
A recursive $3-\sigma$ rejection scheme was used to suppress outliers. 
Any direct comparison between And V and the globular cluster data must also 
adjust for differences in distance and reddening. While the 
globular cluster distances can be computed using the brightness of 
the HB, the only standard candle currently available for And V is the RGB-tip, 
and the use of different standard candles when comparing the globular cluster 
and And V sequences is a potential source of 
systematic error\footnote[2]{Why not compute distances for the clusters 
using the RGB-tip? We were reluctant to do this because (1) the NGC 288 
data does not sample the RGB-tip, (2) in some systems the brightness of the 
RGB-tip can be affected by dynamical evolution (e.g.\ Davidge 1995; 2000) and 
(3) the number density of stars near the RGB-tip can be small, and this 
can bias RGB-tip measurements (Davidge 2001).}. 
Therefore, we adopted calibrations for the RGB-tip and HB that give 
consistent distance moduli. 

	Salaris \& Cassisi (1997) investigate the RGB-tip and HB distance 
scales using calibrations based on stellar evolution models, 
and the results from that study can be used to 
establish consistent distance scales for And V and the globular clusters. 
An initial distance modulus to And V was computed using the Armandroff et al.\ 
(1998) $I-$band RGB-tip measurement and Equation 9 of Salaris \& 
Cassisi (1997). We elected to use the Armandroff et al.\ (1998) RGB-tip 
measurement because it is in the same photometric system as the calibration. 
Nevertheless, we note that the RGB-tip occurs at $i' = 20.95$ in our data, 
which corresponds roughly to $I = 20.8$; hence, the RGB-tip brightness 
measured here is consistent with the Armandroff et al.\ (1998) value of $I = 
20.85$. The RGB-tip calibration is mildly sensitive to metallicity, with 
M$_{I}^{RGB-tip}$ differing by 0.09 mag between [Fe/H] = $-1.6$ and $-2.2$. 
Given this difference, we computed the distance modulus 
of And V using both [Fe/H] $=-1.6$ and $-2.2$, and 
found that the comparisons in Figure 5, described below, are 
insensitive to the adopted metallicity. Cluster distance moduli were calculated 
using the HB $V$ magnitudes listed in the current version of the Harris (1996) 
database and Equation 3 of Salaris \& Cassisi (1997), 
with ZAHB corrections from Equation 5 of Cassisi \& Salaris (1997). 

	The RGB-tip and HB calibrations might be expected to give different 
distance moduli because of uncertainties in the models of stars approaching 
the core He flash and experiencing core He burning. In fact, 
Salaris \& Cassisi (1997) compared the distance moduli of seven 
galaxies computed using RGB-tip and HB measurements, and the data in their 
Table 4 indicate that there is a $\sim 0.15$ mag 
offset, in the sense that the distance moduli computed with the HB tend to be 
lower. The entries for NGC 205 in Table 4 of Salaris \& Cassisi (1997) were not 
included when computing this mean because the difference between the HB and 
RGB-tip distances for this galaxy is clearly discrepant. To bring the 
zeropoints of the RGB-tip and HB calibrations into agreement we subtracted 
0.15 mag from the RGB-tip distance modulus of And V.

	The standard candle brightness, reddening, 
and final distance modulus of And V and the clusters 
are listed in Table 2. The reddening for And V listed in Table 2 is from 
Burstein \& Heiles (1982), while $E(B-V)$ values for the clusters are 
from the current version of the Harris (1996) database. 
The reddening for And V listed in Table 2 is not greatly different 
from that predicted by the Schlegel, Finkbeiner, \& Davis (1998) extinction 
maps, which give $E(B-V) = 0.13$ for And V\@. $E(g'-i')$ was computed from 
$E(B-V)$ using the entries listed in Table 6 of Schlegel et al.\ 
(1998). The distance scale is a matter of active 
debate, and we appreciate that other calibrations are available for the 
HB and RGB-tip; however, we are confident that the distances in 
Table 2 computed from the HB and RGB-tip are consistent, which is of 
fundamental importance for the comparisons described below.

	The globular cluster RGB sequences are compared 
with the And V inner annulus data in the top panel of Figure 5. 
It is reassuring that the RGB-tip brightnesses 
of And V, NGC 2419, and M2 agree, confirming that the adopted 
RGB-tip and HB calibrations are consistent. A visual inspection of the top 
panel of Figure 5 indicates that the And V 
data closely matches the RGB of NGC 2419, although when 
M$_{i'}$ is between --2 and --3 there may be a tendency for the And V 
data to fall midway between the NGC 2419 and M2 sequences. This comparison 
suggests that the majority of stars in And V have a metallicity comparable to 
NGC 2419, for which [Fe/H] $= -2.1$. 

	Uncertainties in the photometric calibration and the reddening 
complicate comparisons like those in the top panel of Figure 5. For example, 
if $E(B-V) = 0.13$ is adopted for And V, as predicted from the Schlegel et 
al.\ (1998) maps, then the And V sequence falls midway between the NGC 2419 
and M2 giant branches. The shape of the upper RGB 
provides a means of estimating metallicity that is insensitive to 
uncertainties in the photometric calibration and reddening (e.g. Hartwick \& 
Sandage 1968; Sarajedini 1994), and so we use the slope 
of the upper giant branch to set constraints on the 
metallicity of And V. The giant branch of And V is steeper than that of 
M2, and this is demonstated in the bottom panel of Figure 5, 
where the And V data from the top panel has been shifted by 0.07 mag 
along the color axis to force agreement with the M2 RGB when M$_{i'} > -3$. 
When compared in this manner, stars with M$_{i'} < -3$ in And V fall to 
the left of the M2 sequence, even though the And V data matches the 
M2 giant branch when M$_{i'} > -3$. Thus, the And V 
RGB is steeper than that of M2, which is consistent with And V being 
more metal-poor than this cluster.

	In an effort to compare the upper RGB slopes in a quantifiable manner, 
a least squares fit was made to the upper RGBs of And V, NGC 2419, and M2 to 
determine $\frac{\Delta(g'-i')}{\Delta(i')}$, and the results are listed in 
Table 3. NGC 288 was not considered, as our observations do not sample the 
upper RGB of this cluster. The slopes listed in Table 3 were measured from 
stars with M$_{i'}$ between $-3$ and $-4$, although the results are 
insensitive to the magnitude interval over which the slopes are calculated; for 
example, the conclusions do not change if slopes are measured using stars with 
M$_{i'}$ between --2 and --4. Obvious AGB and field stars were excluded from 
the analysis.

	The $\frac{\Delta(g'-i')}{\Delta(i')}$ values for NGC 2419 and M2 are 
significantly different, in the sense that the RGB of NGC 2419 is steeper than 
that of M2. However, the entries for And V and NGC 2419 in Table 3 
agree within the estimated uncertainties. Using 
observations of globular clusters spanning a range 
of metallicities, Sarajedini (1994) found that 
$\Delta V_{1.2}$, which is a measure of RGB slope based on the difference in 
$V$ between the HB and the RGB at $(V-I)_0 = 1.2$, changes linearly 
with [Fe/H]. If it is assumed that $\frac{\Delta(g'-i')}{\Delta(i')}$ 
also changes in a linear manner with [Fe/H] then 
we find that [Fe/H] $= -2.2 \pm 0.1$ for And V\@. 
We adopt this as our `best' photometric metallicity estimate for And V\@.

\subsection{ The RGB LF of And V and a Search for Luminous AGB Stars}

	The LFs of And V and NGC 2419 are compared in the upper panel of 
Figure 6. The And V LF was constructed using stars within 89 arcsec of the 
galaxy center, with tight limits being applied about the giant 
branch on the CMD to reduce field star contamination. The 
NGC 2419 LF in Figure 6, which was scaled along the vertical axis to match the 
number of stars in the And V LF when M$_{i'} < -1$, is in 
excellent agreement with the And V LF. The good agreement between 
the LFs of And V and NGC 2419 is not unexpected, given that 
the rate of evolution on the RGB does not change markedly with 
metallicity (e.g.\ VandenBerg 1992). Indeed, the $i'$ LFs of NGC 288 ([Fe/H] 
$\sim -1.2$) and M2 ([Fe/H] $\sim -1.6$), which are compared in the lower 
panel of Figure 6, are also very similar.

	We have also searched And V for bright AGB 
stars, which are a signature of star formation during intermediate epochs. The 
dwarf spheroidal companions of the Milky-Way and M31 span a range of 
star-forming histories, with some containing uniformly old populations, while 
others contain populations that formed during intermediate epochs (e.g.\ Mateo 
1998; Da Costa et al.\ 2000). There are indications that the dwarf spheroidal 
systems of the Milky-Way and M31 may have evolved differently, since 
a significant fraction of the Galactic dwarf spheroidals contain 
intermediate-age populations, while the only M31 dwarf spheroidal to show 
such a population is And II (Da Costa et al.\ 2000). The detection 
of an extended AGB in And V would suggest that the M31 and Milky-Way 
systems may not be as different as currently appears to be the case.

	The ($I, V-I$) CMDs of systems with intermediate-age populations (e.g.\ 
Figure 5 of Richer, Crabtree, \& Pritchet 1990), suggest that if And V has an 
extended AGB then it will likely have a peak $i'$ brightness within $\sim 1.0$ 
mag of the RGB-tip, and extending $\leq 1$ mag in $(g'-i')$ redward of the 
RGB-tip. The $(i', g'-i')$ CMDs of stars in And V near the RGB-tip in the three 
radial intervals defined in Figure 3 are compared in Figure 7. Foreground stars 
and background galaxies have magnitudes and colors that are similar to 
candidate AGB stars, and objects were counted in 0.5 mag ($i') \times 1.0$ mag 
($g'-i'$) regions of each CMD to assess statistically this contamination; if 
And V contains a suitably large population of bright AGB stars then they will 
show up as an excess in the innermost annulus.

	The boxes used for the star counts, and the number 
of objects detected in each, are indicated in Figure 7. The mean number of 
stars in each box is 5.3, with a standard deviation $\pm 2.7$. Thus, the 9 
objects in the innermost annulus with $i'$ between 20.5 and 
21.0, which is where an AGB component might be expected, do not constitute a 
statistically significant excess population. 
This does not mean the And V lacks an intermediate-age population, 
but simply that it is not large enough to produce a statistically 
significant number of bright AGB stars. 
Spectra and near-infrared photometry will be useful to determine 
if any bright sources near the center of And V are main sequence foreground 
stars, giants at the distance of And V, or background objects. We note 
that since And V is very metal-poor, then AGB-tip stars belonging to the galaxy 
may be Carbon stars, which can be identified using narrow-band imaging 
techniques (e.g.\ Richer, Pritchet, \& Crabtree 1985; Cook, Aaronson, 
\& Norris 1986).

\section{DISCUSSION AND CONCLUSIONS}

	Deep $g'$, $r'$, and $i'$ images obtained with GMOS on the Gemini North 
telescope have been used to investigate the photometric properties of bright 
giants in the dwarf spheroidal galaxy And V\@. The data include 
stars that are three magnitudes in $i'$ fainter than the RGB-tip, making the 
current paper the deepest photometric study of this galaxy to date. Based on 
the slope of the upper RGB on the $(i', g'-i')$ CMD, we conclude that And V has 
a metallicity [Fe/H] $= -2.2 \pm 0.1$, which is significantly lower than was 
estimated by Armandroff et al.\ (1998). With the previous metallicity estimate 
of [Fe/H] $= -1.5$, And V had the largest deviation from the trend between 
[Fe/H] and M$_V$ among dwarf galaxies plotted in Figure 4 of Caldwell (1999); 
however, our revised metallicity places And V squarely on this relation.

	A number of factors, including environment, 
likely contribute scatter to the relation between M$_V$ and [Fe/H], 
and studies such as ours provide data that will ultimately allow 
the intrinsic dispersion about the [Fe/H] -- M$_V$ 
relation to be measured. Our revised metallicity for And V tightens the 
relation between integrated brightness and metallicity for low 
mass systems, and one implication of such a tight relation is that the M/L 
ratio of dwarf systems must not differ by large amounts at a given M$_V$.
The chemical contents of dwarf elliptical galaxies in 
the M81 group appear to be defined by total baryonic mass (Caldwell 
et al.\ 1998), rather than other properties such as the central stellar 
concentration, possibly suggesting that the [Fe/H] -- M$_V$ relation has its 
origins in global galaxy properties. A caveat is that the structural properties 
of low mass galaxies in hierarchal systems may change with time; 
for example, tidal effects may alter the structural properties of dwarf 
systems (e.g.\ Cuddeford \& Miller 1990, Mayer et al.\ 2001).

	There is an HI cloud close to And V on the 
sky but, with $v_{\odot} = -176 \pm 1$ km/sec 
(Blitz \& Robishaw 2000), the cloud has a velocity that is very different 
from that of And V, for which Evans et al.\ (2000) measure $v_{\odot} = -403 
\pm 4$ km/sec, indicating that the cloud is likely a chance superposition. 
The absence of an HI reservoir for And V may not be surprising 
since, with a distance of 118 kpc from the center 
of M31, this galaxy falls within the 250 kpc radius where Local Group 
dwarf spheroidals appear to have low HI contents (Blitz \& Robishaw 2000). 
In fact, we do not find a statistically signicant population of upper AGB stars 
near the center of And V, suggesting that the galaxy did not experience 
significant star formation during intermediate epochs. At 
present, the only dwarf spheroidal companion of M31 to show evidence for star 
formation during intermediate epochs is And II (Da Costa et al.\ 2000), 
whereas many of the dwarf spheroidal companions of the Milky-Way contain 
intermediate-age populations. It thus appears that the star-forming histories 
of the Milky-Way and M31 dwarf spheroidal systems may have been systematically 
different. 

\acknowledgements

It is a pleasure to thank the Project Scientists (David Crampton and 
Roger Davies), the Project Manager (Rick Murowinski), and the countless 
others at the Herzberg Institute of Astrophysics, the Astronomy Technology 
Center, Durham University, and the Gemini Observatory who worked so hard to 
design, construct, deliver and commission GMOS. The authors also thank the 
referee for detailed comments on the first version of this paper.

\clearpage

\begin{table*}
\begin{center}
\begin{tabular}{lccc}
\tableline\tableline
Cluster & Filter & Exposure Time & FWHM \\
 & & (Sec) & \\
\tableline
And V & $g'$ & $9 \times 150$ & $0\farcs58$ \\
 & $r'$ & $9 \times 100$ & $0\farcs80$ \\
 & $i'$ & $9 \times 100$ & $0\farcs58$ \\
NGC 288 & $g'$ & $4 \times 6$ & $0\farcs88$ \\
 & $i'$ & $4 \times 3$ & $0\farcs73$ \\
NGC 2419 & $g'$ & $4 \times 90$ & $0\farcs58$ \\
 & $r'$ & $4 \times 45$ & $0\farcs58$ \\
 & $i'$ & $4 \times 45$ & $0\farcs58$ \\
M2 (NGC 7089) & $g'$ & $4 \times 6$ & $0\farcs95$ \\
 & $i'$ & $4 \times 3$ & $1\farcs09$ \\
\tableline
\end{tabular}
\end{center}
\caption{Exposure Times and Final Image Qualities}
\end{table*}

\clearpage

\begin{table*}
\begin{center}
\begin{tabular}{lccccc}
\tableline\tableline
Object & [Fe/H] & $I_{RGB-tip}$ & V$_{HB}$ & $E(B-V)$ & (m--M)$_0$ \\
\tableline
And V & -- & 20.85 & -- & 0.16 & 24.8 \\
NGC 288 & --1.24 & -- & 15.30 & 0.03 & 14.57 \\
NGC 2419 & --2.12 & -- & 20.45 & 0.11 & 19.59 \\
M2 & --1.62 & -- & 16.05 & 0.06 & 15.29 \\
\tableline
\end{tabular}
\end{center}
\caption{Standard Candle Brightnesses, Reddenings, and Distance Moduli}
\end{table*}

\clearpage

\begin{table*}
\begin{center}
\begin{tabular}{lc}
\tableline\tableline
Object & $\frac{\Delta(g'-i')}{\Delta(i')}$ \\
\tableline
And V & $-0.340 \pm 0.021$ \\
NGC 2419 & $-0.360 \pm 0.024$ \\
M2 & $-0.494 \pm 0.033$ \\
\tableline
\end{tabular}
\end{center}
\caption{RGB slopes measured from stars with M$_{i'}$ between --3 and --4}
\end{table*}

\clearpage

\begin{center}
FIGURE CAPTIONS
\end{center}

\figcaption
[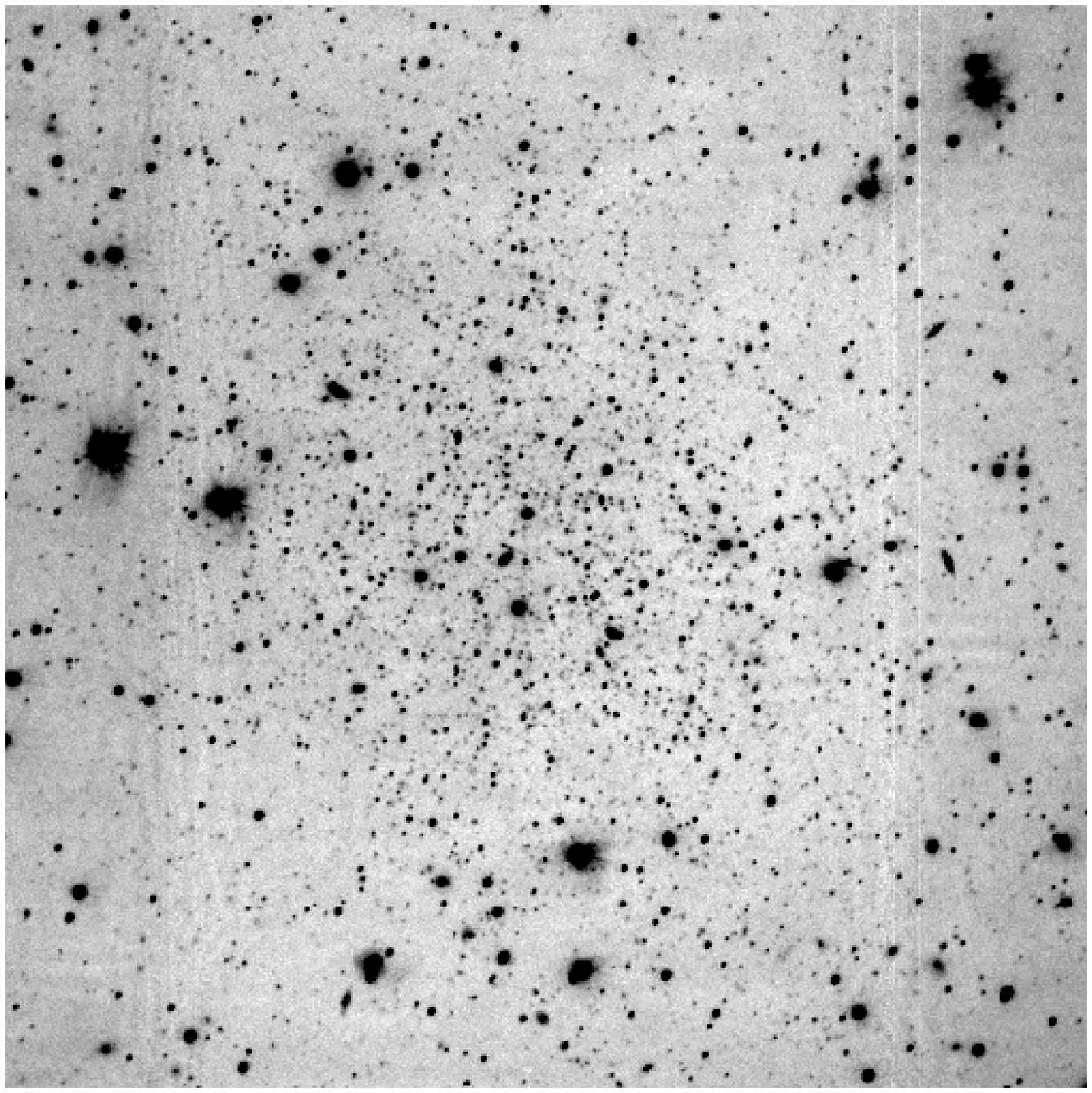]
{The central $110\arcsec \times 110\arcsec$ of the final $i'$ image. North 
is at the top, and East is to the left. The main body of And V is clearly 
evident near the field center.}

\figcaption
[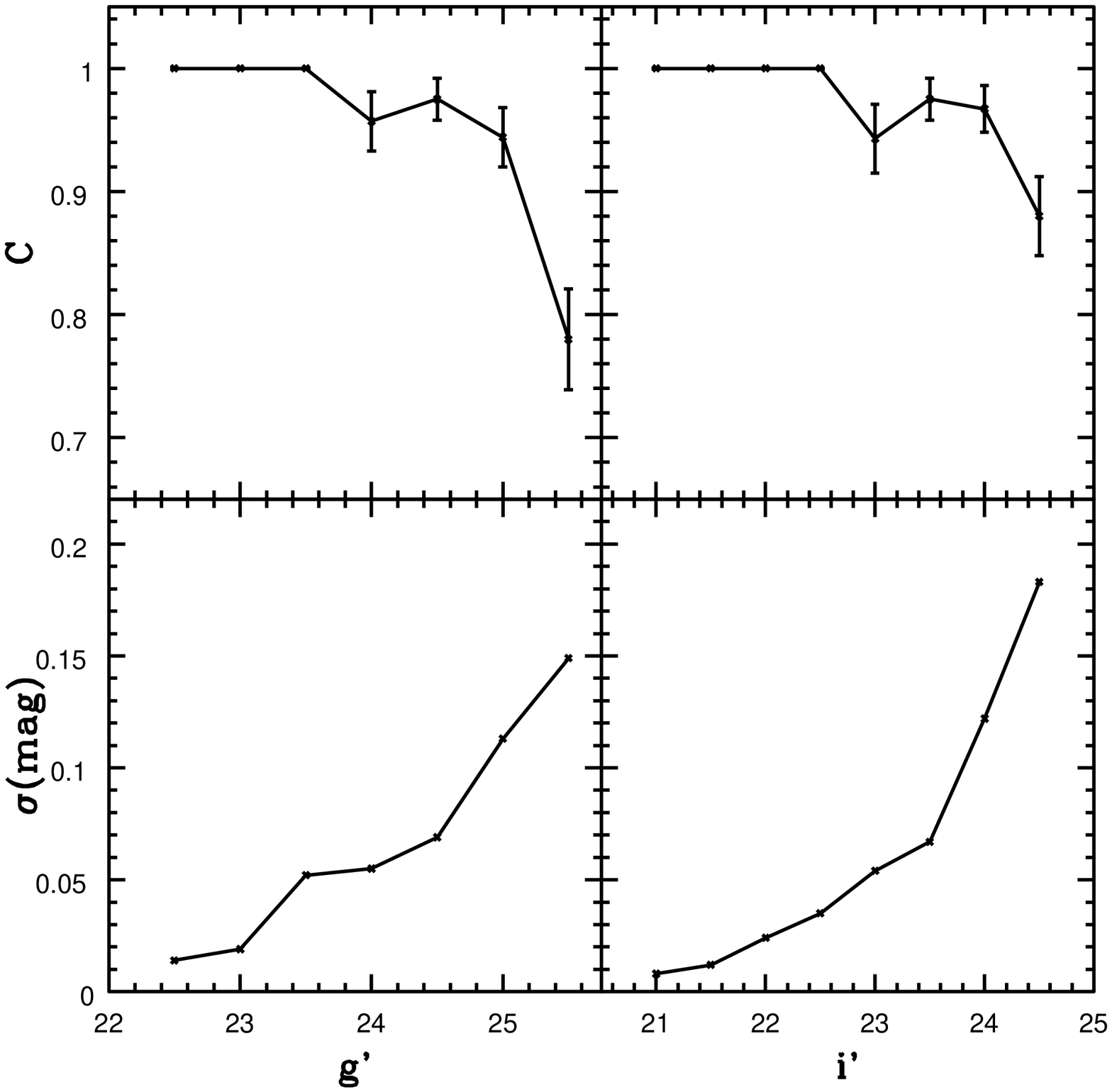]
{The completeness fractions and estimated internal photometric uncertainties 
in $g'$ and $i'$, as predicted from the artificial star experiments. $C$ 
is the ratio of recovered to added stars, while $\sigma$ 
is the standard deviation of the difference between the measured and actual 
stellar brightness.}

\figcaption
[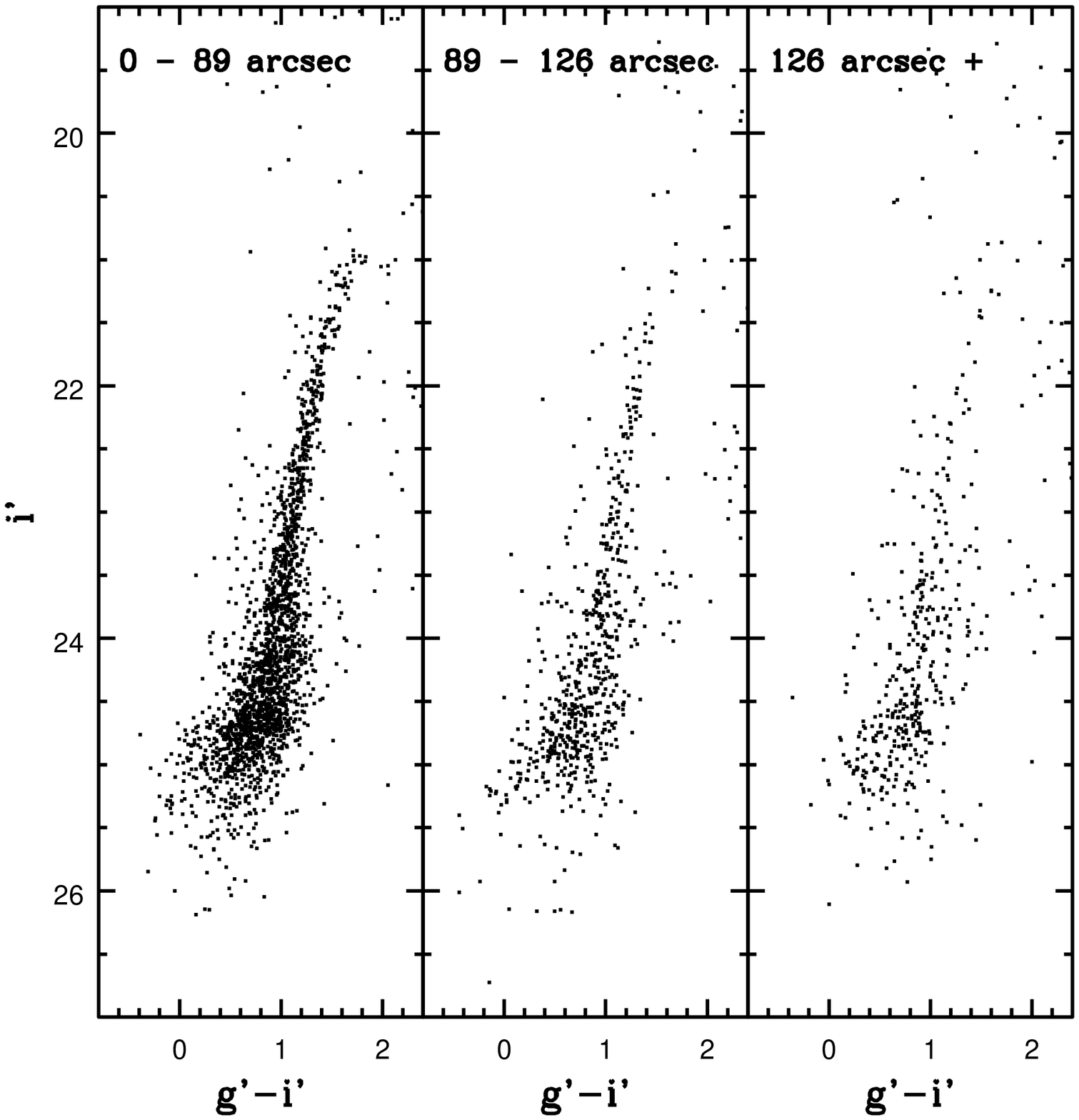]
{The $(i', g'-i')$ CMDs of stars in Figure 1. The radial intervals 
specified near the top of each panel are measured from the center of And V\@.
Note that the And V giant branch is seen even in the outermost annulus.} 

\figcaption
[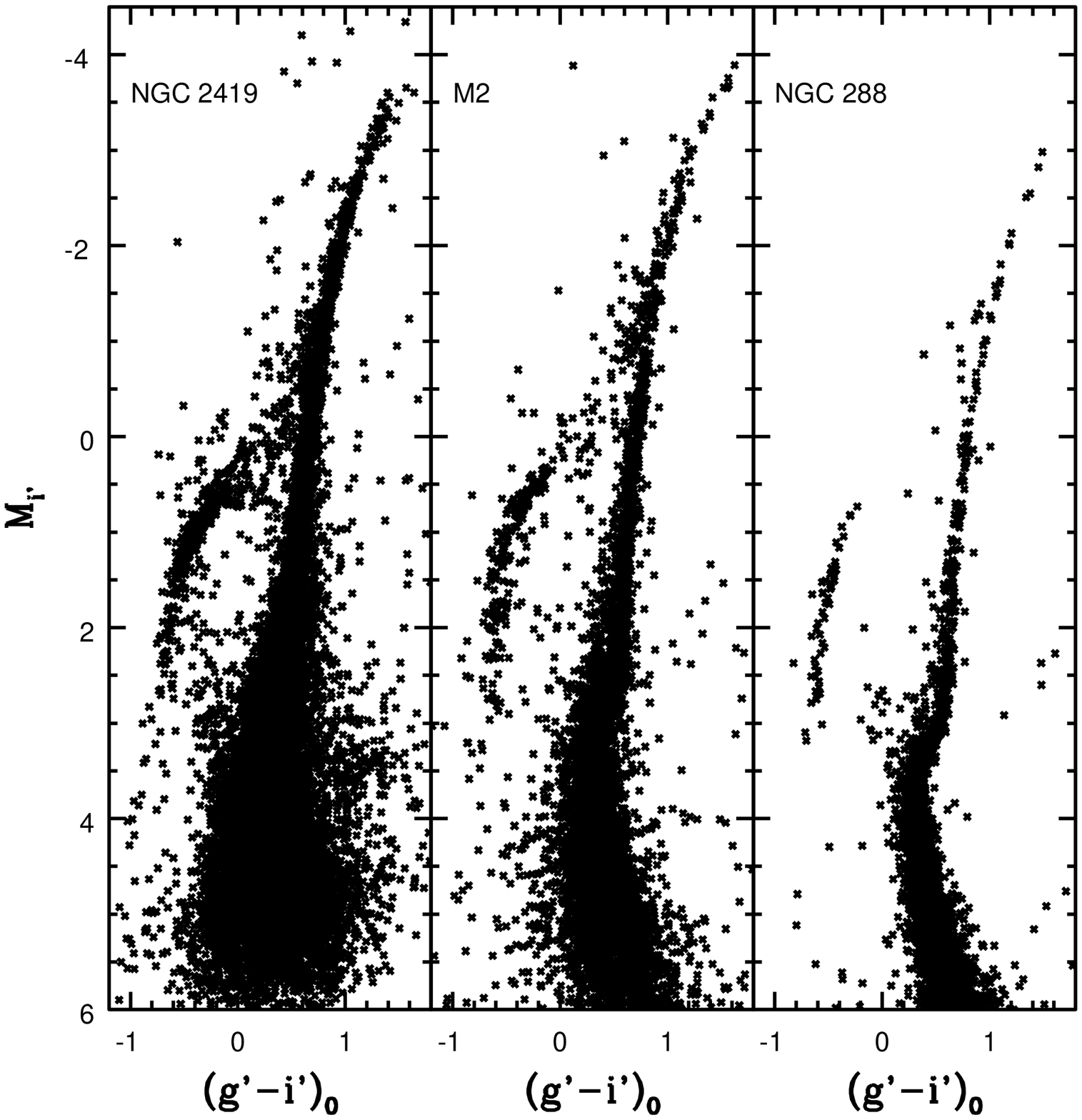]
{The CMDs of the clusters NGC 2419, M2, and NGC 288. The CMDs have been 
corrected for distance and reddening using the procedure described 
in the text.}

\figcaption
[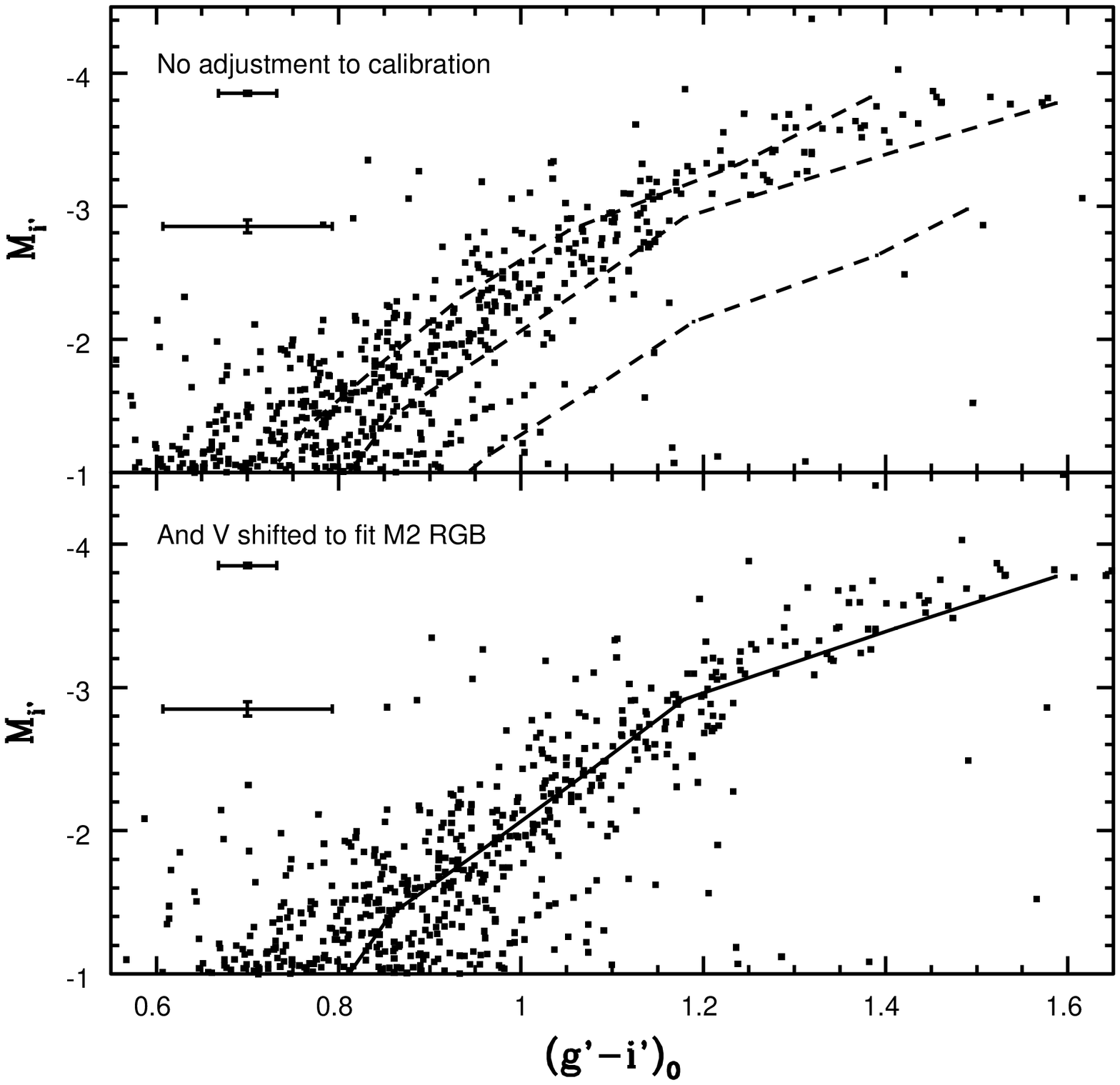]
{The annulus 1 CMD of And V is compared with the normal point sequences for 
the globular clusters NGC 2419, M2, and NGC 288, which are plotted as dashed 
lines, in the top panel. The NGC 288 sequence terminates $\sim 1$ mag fainter 
than the other clusters because our data do not sample the upper 
RGB of this cluster. In the bottom panel, the 
And V data have been shifted by 0.07 mag along the color axis to force 
agreement with the M2 RGB, shown as a solid line, 
when M$_{i'} > -3$. Note that the brightest giants 
in And V fall consistently blueward of the M2 sequence in the lower panel, 
indicating that And V is more metal-poor than M2.}

\figcaption
[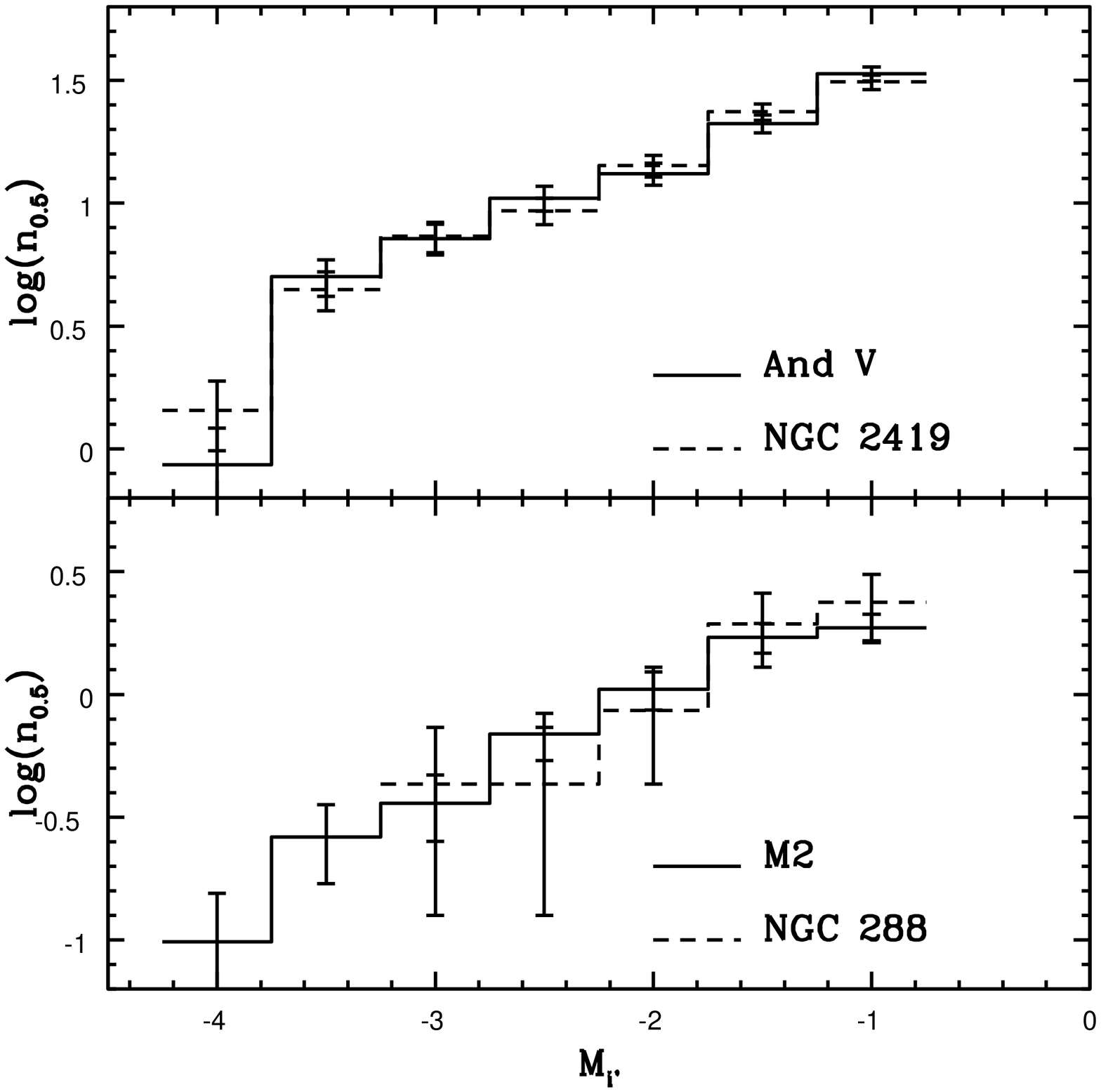]
{The M$_{i'}$ LFs of And V and NGC 2419 (top panel) and 
M2 and NGC 288 (lower panels). n$_{05}$ is the number of stars per 
0.5 mag interval in M$_{i'}$ per square arcmin, and the error bars show the 
uncertainties due to counting statistics. The NGC 2419 and NGC 288 LFs have 
been scaled to match the number of stars with M$_{i'} < -1$ in And V and M2, 
respectively. Note the excellent agreement between each pair of LFs.}

\figcaption
[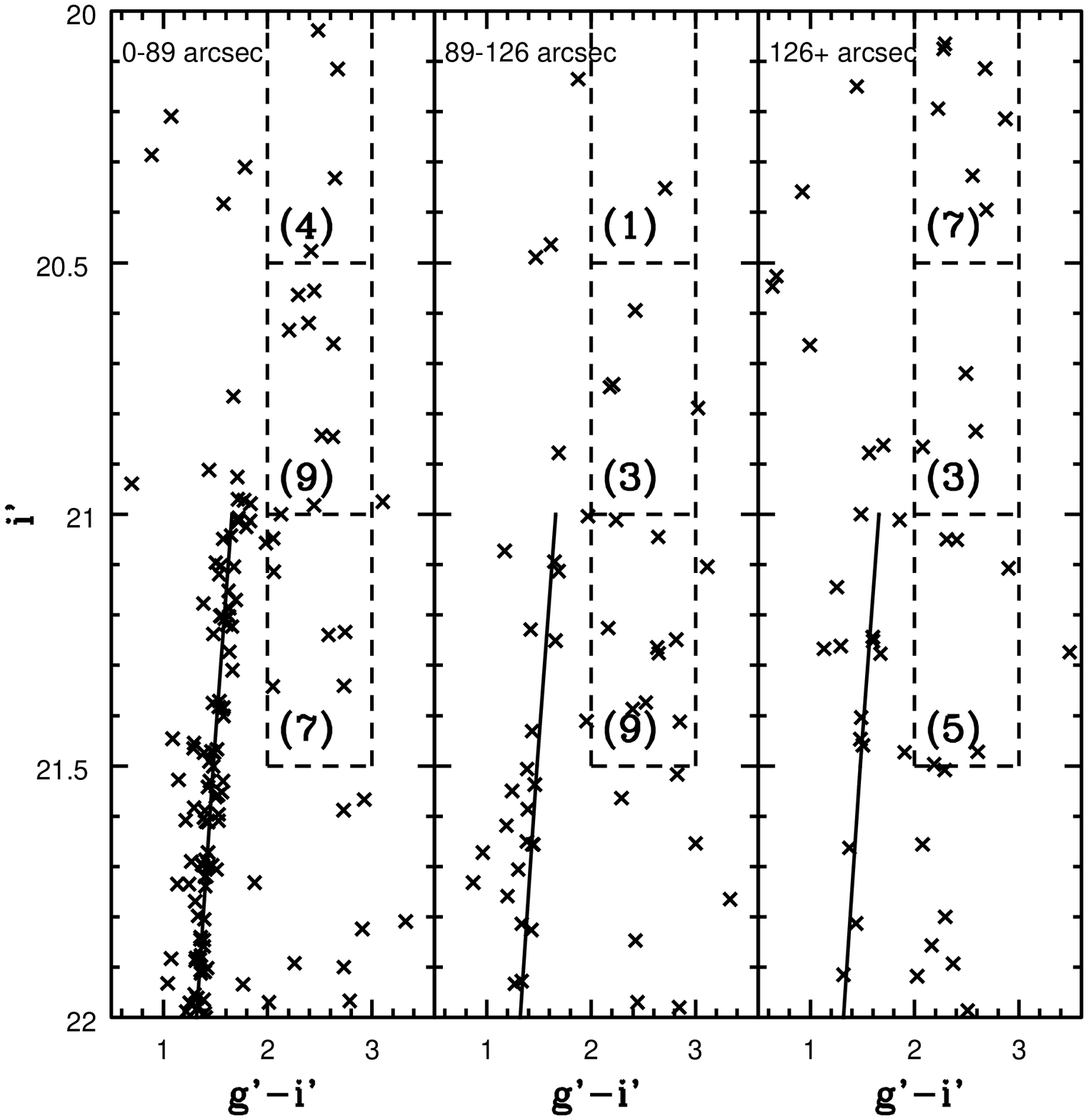]
{The $(i', g'-i')$ CMDs of stars in three annuli, centered on 
And V and sampling comparable areas on the sky. 
The solid line shows the giant branch locus of And V defined by 
normal points, while the dashed lines delineate regions used to search 
for bright AGB stars; the number of objects in each box is given in brackets. 
The number of objects detected in the innermost annulus between $i' = 20.5$ and 
21.0, which is where bright AGB stars might be expected, 
does not differ significantly from the number of forground and background 
objects.}


\clearpage

\begin{references}

\reference{}Armandroff, T. E., \& Da Costa, G. S. 1991, AJ, 101, 1329

\reference{}Armandroff, T. E., Davies, J. E., \& Jacoby, G. H. 1998, AJ, 116, 2287

\reference{}Blitz, L., \& Robishaw, T. 2000, ApJ, 541, 675

\reference{}Bullock, J. S., Kravtsov, A. V., \& Weinberg, D. H. 2001, ApJ, 548, 33

\reference{}Burstein, D., \& Heiles, C. 1982, AJ, 87, 1165

\reference{}Caldwell, N. 1999, AJ, 118, 1230

\reference{}Caldwell, N., Armandroff, T. E., Da Costa, G. S., \& Seitzer, P. 1998, AJ, 115, 535

\reference{}Cassisi, S., \& Salaris, M. 1997, MNRAS, 285, 593

\reference{}Cole, S., Lacey, C. G., Baugh, C. M., \& Frenk, C. S. 2000, MNRAS, 319, 168

\reference{}Cook, K. H., Aaronson, M., \& Norris, J. 1986, 305, 634

\reference{}Crampton, D. et al.\ 2000, SPIE, 4008, 114

\reference{}Cuddeford, P., \& Miller, J. C. 1990, MNRAS, 244, 64

\reference{}Da Costa, G. S., Armandroff, T. E., \& Caldwell, N. 2002, AJ, 124, in press.

\reference{}Da Costa, G. S., Armandroff, T. E., Caldwell, N., \& Seitzer, P. 2000, AJ, 119, 705

\reference{}Davidge, T. J. 1995, AJ, 110, 1177

\reference{}Davidge, T. J. 2000, AJ, 120, 1853

\reference{}Davidge, T. J. 2001, AJ, 121, 3100

\reference{}Davies, R. L. et al.\ 1997, SPIE, 2871, 1099

\reference{}Dekel, A., \& Silk, J. 1986, ApJ, 303, 39

\reference{}Evans, N. W., Wilkinson, M. I., Guhathakurta, P., Grebel, E. K., \& Vogt, S. 2000, ApJ, 540, L9

\reference{}Forbes, D. A., Masters, K. L., Minniti, D., \& Barmby, P. 2000, A\&A, 358, 471

\reference{}Fukugita, M., Ichikawa, T., Gunn, J. E., Doi, M., Shimasaku, K., \& Schneider, D. P. 1996, AJ, 111, 1748

\reference{}Harris, W. E. 1996, AJ, 112, 1487

\reference{}Hartwick, F. D. A., \& Sandage, A. 1968, ApJ, 153, 715

\reference{}Ibata, R., Irwin, M., Lewis, G. F., \& Stolte, A. 2001, ApJ, 547, L133

\reference{}Kauffmann, G., White, S. D. M., \& Guiderdoni, B. 1993, MNRAS, 264, 201

\reference{}Klypin, A., Kravtsov, A. V., Octavio, V., \& Francisco, P. 1999, ApJ, 522, 82

\reference{}Kobayashi, C., \& Arimoto, N. 1999, ApJ, 527, 573

\reference{}Kodama, T., \& Arimoto, N. 1997, A\&A, 320, 41

\reference{}Landolt, A. U. 1992, AJ, 104, 340

\reference{}Mart\'{i}nez-Delgado, D., Aparicio, A., G\'{o}mez-Flechoso, M. A., \& Carrera, R. 2001, ApJ, 549, L199

\reference{}Mateo, M. 1998, ARAA, 36, 435

\reference{}Mayer, L., Governato, F., Colpi, M., Moore, B., Quinn, T., Wadsley, J., Stadel, J., \& Lake, G. 2001, ApJ, 547, L123

\reference{}Richer, H. B., Pritchet, C. J., \& Crabtree, D. R. 1985, ApJ, 298, 240

\reference{}Richer, H. B., Crabtree, D. R., \& Pritchet, C. J. 1990, ApJ, 355, 448

\reference{}Salaris, M., \& Cassisi, S. 1997, MNRAS, 289, 406

\reference{}Sarajedini, A. 1994, AJ, 107, 618

\reference{}Saviane, I., Held, E. V., \& Bertelli, G. 2000, A\&A, 355, 56

\reference{}Schlegel, D. J., Finkbeiner, D. P., \& Davis, M. 1998, ApJ, 500, 525

\reference{}Shetrone, M. D., C\^{o}t\'{e}, P., \& Sargent, W. L. W. 2001, ApJ, 548, 592

\reference{}Somerville, R. S., \& Primack, J. R. 1999, MNRAS, 310, 1087
 
\reference{}Stetson, P. B. 1987, PASP, 99, 191

\reference{}Stetson, P. B., \& Harris, W. E. 1988, AJ, 96, 909

\reference{}Trager, S. C., King, I. R., \& Djorgovski, S. 1995, AJ, 109, 218

\reference{}VandenBerg, D. A., 1992, ApJ, 391, 685

\end{references}
\end{document}